\documentstyle[fleqn,amssymb]{elsart}

\newcommand{\cB}{{\cal B}}

\newcommand{\cE}{{\cal E}}
\newcommand{\cF}{{\cal F}}

\newcommand{\cH}{{\cal H}}

\newcommand{\cK}{{\cal K}}

\newcommand{\cS}{{\cal S}}

\newcommand{\defeq}{:=}

\newcommand{\set}[1]{\{ #1 \}}

\newcommand{\map}[3]{#1 : #2 \rightarrow #3}
\newcommand{\maps}[2]{#1 \mapsto #2}
\def\id {{\rm id}}

\newcommand{\abs}[1]{\left| #1 \right|}
\newcommand{\norm}[1]{\left \| #1 \right \|}
\newcommand{\trnorm}[1]{\left \| #1 \right \|_1}
\newcommand{\cbnorm}[1]{\left \| #1 \right \|_{\rm cb}}

\newcommand{\tensprod}[2]{#1 \otimes #2}
\newcommand{\tr}[1]{{\rm tr} #1}
\newcommand{\ptr}[2]{{\rm tr}_{#1} #2} 

\newcommand{\rp}[1]{{\rm Re}\, #1}

\newcommand{\trn}[1]{#1 ^\top}
\def\su {\rm SU}

\def\idty{{\leavevmode{\rm 1\mkern -5.4mu I}}}

\newcommand{\ket}[1]{| #1 \rangle}

\newcommand{\braket}[2]{\langle #1 | #2 \rangle}
\newcommand{\ketbra}[2]{| #1 \rangle \langle #2 | }

\def\drangle{{\leavevmode{\rm \rangle\mkern -4mu \rangle}}}
\def\dlangle{{\leavevmode{\rm \langle\mkern -4mu \langle}}}

\newcommand{\dket}[1]{| #1 \drangle}

\newcommand{\dbraket}[2]{\dlangle #1 | #2 \drangle}
\newcommand{\dketbra}[2]{| #1 \drangle \dlangle #2 |}

\def \acin{{Ac\' \i n }}

\begin{document}

\begin{frontmatter}

\title{A fidelity measure for quantum channels}

\author{Maxim Raginsky}

\address{Center for Photonic Communication and Computing\\
Department of Electrical and Computer Engineering\\
Northwestern University, Evanston, Illinois 60208-3118\\
E-mail address:  {\tt maxim@northwestern.edu}}

\begin{abstract}
We propose a fidelity measure for quantum channels in a straightforward analogy to the corresponding mixed-state fidelity of Jozsa. We describe properties of this fidelity measure and discuss some applications of it to quantum information science.


\end{abstract}

\end{frontmatter}

\section{Introduction}

Given any pair $\rho,\sigma$ of density operators, the fidelity \cite{jozsa}
\begin{equation}
F(\rho,\sigma) \defeq
	\left( 
		\tr{\sqrt{\sqrt{\rho}{\sigma}\sqrt{\rho}}}
	\right)^2
\label{eq:statefid}
\end{equation}
quantifies the extent to which $\rho$ and $\sigma$ can be distinguished from one another.  The properties it possesses are quite natural from the physical point of view.  For instance, $0 \le F(\rho,\sigma) \le 1$, and $F(\rho,\sigma)=1$ if and only if $\rho = \sigma$; for any unitary transformation\footnote{In this paper, given an operator $X$, the adjoint operator will be denoted by $X^*$, not $X^\dag$.} $U$, $F(U \rho U^*, U \sigma U^*) = F(\rho,\sigma)$; and, according to a theorem of Uhlmann \cite{jozsa,uhlmann1},
\begin{equation}
F(\rho,\sigma) = \max_{\ket{\psi_\rho},\ket{\psi_\sigma}} \abs{\braket{\psi_\rho}{\psi_\sigma}}^2,
\label{eq:uhlth}
\end{equation}
where $\ket{\psi_\rho}$ and $\ket{\psi_\sigma}$ are purifications of $\rho$ and $\sigma$ on an extended Hilbert space.  Furthermore, the fidelity (\ref{eq:statefid}) is known to be equivalent to other measures of distinguishability of quantum states \cite{fvdg}.

In a paper of Childs et al. \cite{cpr} it was suggested that a notion of distinguishability of quantum channels would be as vital to quantum information science as the corresponding notion for quantum states.  [We recall that a quantum channel between two systems with Hilbert spaces $\cH$ and $\cK$ respectively is a completely positive trace-preserving linear map from the set $\cS(\cH)$ of density operators on $\cH$ to the set $\cS(\cK)$ of density operators on $\cK$].  In this letter we show that a fidelity measure for a pair of quantum channels $S$ and $T$ can be derived from the fidelity (\ref{eq:statefid}) for suitably chosen density operators $\rho_S$ and $\rho_T$, and that the proposed fidelity measure enjoys properties that are quite similar to those of the mixed-state fidelity.  We also prove an analogue of Uhlmann's theorem, formulated in terms of dilations of channels on an extended Hilbert space.  We then discuss our fidelity measure in the context of (a) distinguishing channels via superdense coding, (b) improving distinguishability of channels through preprocessing, and (c) characterizing the performance of quantum error-correcting codes.

\section{Duality between quantum channels and bipartite states}

In order to state our results in clear and compact form, we briefly recall a very convenient characterization of quantum channels by means of bipartite states.  Let $T$ be a channel that converts systems with the finite-dimensional Hilbert space $\cH$ into systems with the finite-dimensional Hilbert space $\cK$.  It is well-known that the action of $T$ can be described by giving its {\em Kraus operators}, $\map{V_\alpha}{\cH}{\cK}$, so that, for any density operator $\rho \in \cS(\cH)$,
\begin{equation}
T(\rho) = \sum_\alpha V_\alpha \rho V^*_\alpha.
\label{eq:krausrep}
\end{equation}
Since $T$ is trace-preserving, the Kraus operators must satisfy the relation
\begin{equation}
\sum_\alpha V^*_\alpha V_\alpha = \idty_\cH.
\label{eq:krarel}
\end{equation}
The Kraus representation (\ref{eq:krausrep}) is unique only up to unitary equivalence, and thus does not provide a one-to-one characterization of channels.  One such characterization was described in Ref.~\cite{dp1}.  Given a channel $\map{T}{\cS(\cH)}{\cS(\cK)}$ with $\dim \cH = d < \infty$, we define an operator $R_T$ on $\tensprod{\cK}{\cH}$ through
\begin{equation}
R_T \defeq (\tensprod{T}{\id}) 
	\ketbra{\varphi^+_{\cH}}{\varphi^+_{\cH}},
\label{eq:ropdef}
\end{equation}
where "$\id$" denotes the identity channel, and $\ket{\varphi^+_{\cH}} \in \cH^{\otimes 2}$ is the unnormalized maximally entangled state $\sum^d_{i=1}\tensprod{\ket{e_i}}{\ket{e_i}}$.  Then the action of $T$ on an arbitrary density operator $\rho$ on $\cH$ is given by
\begin{equation}
T(\rho) = \ptr{\cH}{[(\tensprod{\idty}{\trn{\rho}})R_T]},
\label{eq:channeldef}
\end{equation}
where $\trn{\rho}$ denotes the transpose of $\rho$ in the basis $\set{\ket{e_i}}$.  We must be careful with Eq.~(\ref{eq:channeldef}) because (a) the subscript $\cH$ refers to the {\em second} factor in the tensor product, since both $\tensprod{\idty}{\trn{\rho}}$ and $R_T$ are operators that act on $\tensprod{\cK}{\cH}$; and (b) all matrix operations (e.g., the transpose) refer to the basis which has been fixed beforehand, when defining $R_T$.  It turns out that the correspondence $\maps{T}{R_T}$ is one-to-one, i.e., $R_S = R_T$ if and only if $S=T$ (see Ref.~\cite{dp1} for a detailed discussion).  The requirement for $T$ to be trace-preserving translates into \cite{dp1}
\begin{equation}
\ptr{\cK}{R_T} = \idty_\cH.
\label{eq:roptp}
\end{equation}

This formalism can be extended to channels acting on bipartite systems \cite{cdkl}.  Let $T$ be a channel that maps states on $\tensprod{\cH_1}{\cH_2}$ to states on $\tensprod{\cK_1}{\cK_2}$ (here we assume, for simplicity, that $\dim \cH_1 = \dim \cH_2 = d$).  Then we define
\begin{equation}
R_T \defeq (\tensprod{T_{\rm odd}}{\id_{\rm even}})
	(\tensprod
		{\ketbra {\varphi^+_{\cH_1}} {\varphi^+_{\cH_1}} }
		{\ketbra {\varphi^+_{\cH_2}} {\varphi^+_{\cH_2}} } ),
\label{eq:ropdefbipart}
\end{equation}
where $\tensprod{T_{\rm odd}}{\id_{\rm even}}$ signifies that the channel $T$ acts only on odd-numbered factors of $\tensprod
		{\ketbra {\varphi^+_{\cH_1}} {\varphi^+_{\cH_1}} }
		{\ketbra {\varphi^+_{\cH_2}} {\varphi^+_{\cH_2}} } $, while the identity channel acts on even-numbered factors.
The operator $R_T$ defined in Eq.~(\ref{eq:ropdefbipart}) acts on $\tensprod {\tensprod{\cK_1}{\cH_1}} {\tensprod{\cK_2}{\cH_2} }$, and the action of the channel $T$ on any $\rho \in \cS(\tensprod{\cH}{\cH_3})$ is given by
\begin{equation}
T(\rho) = \ptr{\cH_1 \cH_2}
	[ (\tensprod
		{\idty_{\tensprod{\cK_1}{\cK_2}}}
		{\trn{\rho}_ {\tensprod{\cH_1}{\cH_2}}}
	) R_T].
\label{eq:channeldefbipart}
\end{equation}

All of the above relations can be cast into a particularly nice form by exploiting the correspondence between an operator $\map{A}{\cH}{\cK}$ and a vector $\dket{A}$ in $\tensprod{\cK}{\cH}$ \cite{royer}.  Fix orthonormal bases $\set{\ket{e_i}}$ and $\set{\ket{e_\mu}}$ of $\cK$ and $\cH$ respectively.  Then the required correspondence is given by
\begin{equation}
A = \sum_{i,\mu} A_{i\mu} \ketbra{e_i}{e_\mu} \quad \longmapsto \quad \dket{A} \defeq \sum_{i,\mu} A_{i\mu} \tensprod{ \ket{e_i}} {\ket{e_\mu}},
\label{eq:liouvcorr}
\end{equation}
so that the double ket $\dket{A}$ denotes the vector in $\tensprod{\cK}{\cH}$ whose components are the matrix elements of $A$. Let $A$ and $B$ be operators acting on $\cK$ and $\cH$ respetively, and let $C$ be an operator from $\cH$ to $\cK$.  Then, using Eq.~(\ref{eq:liouvcorr}), we can readily prove the relations \cite{royer}
\begin{eqnarray}
&& (\tensprod{A}{B}) \dket{C} = \dket {AC\trn{B}}, \label{eq:matrel1} \\
&& \ptr{2}{\dketbra{A}{B}} = AB^*, \label{eq:matrel2} \\
&& \ptr{1}{\dketbra{A}{B}} = \trn{A}\bar{B}, \label{eq:matrel3}
\end{eqnarray}
where the subscripts "1" and "2" refer respectively to the first and second factors in the tensor product, and $\bar{B}$ is the operator whose matrix elements are complex conjugates of the matrix elements of $B$.  Remarkably, the inner product $\dbraket{A}{B}$ is just the Hilbert-Schmidt inner product of the operators $A$ and $B$:  $\dbraket{A}{B} = \tr{(A^*B)}$.

Using the double-ket notation, we can write $\ket{\varphi^+_{\cH}} = \dket{\idty}$ with respect to the basis $\set{\ket{e_i}}$. For any set of Kraus operators $V_\alpha$ for the channel $T$, it can be easily shown that $R_T = \sum_\alpha \dketbra{V_\alpha}{V_\alpha}$.  It is then an immediate consequence of Eqs.~(\ref{eq:matrel1}) and (\ref{eq:matrel2}) that
\begin{equation}
\ptr{\cH}[(\tensprod{\idty}{\trn{\rho}})R_T] = \sum_\alpha \ptr{\cH}\dketbra{V_\alpha \rho}{V_\alpha} = \sum_\alpha V_\alpha \rho V^*_\alpha.
\end{equation}

\section{Fidelity for quantum channels and its properties}

Let $\map{T}{\cS(\cH)}{\cS(\cK)}$ be a channel.  Defining the density operator $\rho_T \defeq (1/d)R_T$, we note that the correspondence $\maps{T}{\rho_T}$ is obviously one-to-one.  Therefore, given two channels $\map{S,T}{\cS(\cH)}{\cS(\cK)}$, we define the corresponding {\em fidelity} as
\begin{equation}
\cF(S,T) \defeq F(\rho_S,\rho_T),
\label{eq:chfid}
\end{equation}
where the right-hand side is the fidelity (\ref{eq:statefid}) between the density operators $\rho_S$ and $\rho_T$.  Properties of the channel fidelity (\ref{eq:chfid}) follow directly from the properties of the mixed-state fidelity (\ref{eq:statefid}) (see Ref.~\cite{jozsa} or Ref.~\cite{nielchu} for the properties of $F$ and the corresponding proofs), and we now list them with proofs.\\

\noindent{{\bf Properties of the channel fidelity} $\cF$}

\begin{enumerate}

\item[{\bf CF1}] $0 \le \cF(S,T) \le 1$, and $\cF(S,T) = 1$ if and only if $S=T$.

\item[{\bf CF2}] $\cF(S,T) = \cF(T,S)$ (symmetry).

\item[{\bf CF3}] For any two unitarily implemented channels $\hat{U}$ and $\hat{V}$ [i.e., $\hat{U}(\rho)=U\rho U^*$ and $\hat{V}(\rho) = V\rho V^*$ with unitary $U$ and $V$], $\cF(\hat{U},\hat{V}) = (1/d^2)\abs{\tr{(U^* V)}}^2$.

\item[{\bf CF4}] For any $0 < \lambda < 1$, $\cF(S,\lambda T_1 + (1-\lambda)T_2) \ge \lambda \cF(S,T_1) + (1-\lambda)\cF(S,T_2)$ (concavity).

\item[{\bf CF5}] $\cF(\tensprod{S_1}{S_2},\tensprod{T_1}{T_2}) = \cF(S_1,T_1)\cF(S_2,T_2)$ (multiplicativity with respect to tensoring).

\item[{\bf CF6}] $\cF$ is invariant under composition with unitarily implemented channels, i.e., for any unitarily implemented channel $\hat{U}$, $\cF(\hat{U} \circ S,\hat{U} \circ T) = \cF(S,T)$.

\item[{\bf CF7}] $\cF$ does not decrease under composition with arbitrary channels, i.e., for any channel $R$, $\cF(R \circ S, R \circ T) \ge \cF(S,T)$.

\end{enumerate}

\noindent{{\bf Proof}}
\begin{enumerate}
\item[{\bf CF1}] follows from the fact that $\rho_S$ and $\rho_T$ are density operators, and from the fact that $\maps{T}{\rho_T}$ is a one-to-one mapping.
\item[{\bf CF2}] --- same reasoning applies.
\item[{\bf CF3}]  $R_{\hat{U}} = \dketbra{U}{U}$, and similarly for $R_{\hat{V}}$. Thus both $\rho_{\hat{U}}$ and $\rho_{\hat{V}}$ are pure states. Since for pure states $\ket{\psi},\ket{\varphi}$ we have $F = \abs{\braket{\psi}{\varphi}}^2$, it follows that $\cF(\hat{U},\hat{V}) = (1/d^2) \abs{\dbraket{U}{V}}^2 =(1/d^2) \abs{\tr{(U^* V)}}^2$.
\item[{\bf CF4}] Note that the map $\maps{T}{\rho_T}$ is linear.  Thus, for $T = \lambda T_1 + (1-\lambda)T_2$, we have $\rho_T = \lambda \rho_{T_1} + (1-\lambda) \rho_{T_2}$, and the concavity of $\cF$ follows from the concavity of the mixed-state fidelity (\ref{eq:statefid}).
\item[{\bf CF5}] It follows from Eq.~(\ref{eq:ropdefbipart}) that $\rho_{\tensprod{S}{T}} = \tensprod{\rho_S}{\rho_T}$, and the multiplicativity property of $\cF$ follows from the corresponding property of $F$.
\item[{\bf CF6}] Write $\tensprod{(\hat{U}\circ T)}{\id} = (\tensprod{\hat{U}}{\id})(\tensprod{T}{\id})$ to obtain $\rho_{\hat{U}T} = (\tensprod{U}{\idty})\rho_T(\tensprod{U^*}{\idty})$, and do the same for $\hat{U} \circ S$.  Since the mixed-state fidelity $F$ is invariant under unitary transformations, the same property holds for the channel fidelity $\cF$.
\item[{\bf CF7}] --- same reasoning as before, except now we have to use the property that $F((\tensprod{R}{\id})\rho_S,(\tensprod{R}{\id})\rho_T) \ge F(\rho_S,\rho_T)$.\hfill $\blacksquare$

\end{enumerate}

Despite its fairly obvious definition, the channel fidelity (\ref{eq:chfid}) possesses properties that are natural from the physical point of view.  First of all, the fact that the channel fidelity is derived from the mixed-state fidelity on a subset of bipartite quantum states implies that, in general, reliable discrimination between quantum channels requires multiparty protocols with entanglement.  Indeed, setting $\rho=\rho_S$ and $\sigma=\rho_T$ in the equation \cite{nielchu}
\begin{equation}
\sqrt{F(\rho,\sigma)} = \min_{\set{F_m}} \sum_m \sqrt{ 
	\tr{(\rho F_m)} \tr{(\sigma F_m)}},
\label{eq:fidpovm}
\end{equation}
where the minimum is taken over all positive operator-valued measures (POVM's) $\set{F_m}$, we see that distinguishability of the channels $S$ and $T$ can be related to distinguishability of probability distributions of outcomes of collective measurements on the bipartite system $\tensprod{\cK}{\cH}$.

Also, property {\bf CF4} implies that mixing channels has adverse effect on distinguishability, and properties {\bf CF6} and {\bf CF7} show that no {\em post}processing, classical or quantum, can render any two channels more distinguishable.  On the other hand, {\em pre}processing may improve distinguishability of channels, as we shall see later.

We mention a couple of useful formulas involving $\cF$.  For instance, the expression for $\cF(T,\id)$, where $T$ is an arbitrary channel, takes particularly simple form.  Let $\set{V_\alpha}$ be any Kraus decomposition of $T$. Then $\rho_T = (1/d)\sum_\alpha \dketbra{V_\alpha}{V_\alpha}$ and, since $\rho_\id \equiv (1/d) \dketbra{\idty}{\idty}$ is pure, we have
\begin{equation}
\cF(T,\id) = F(\rho_T,\rho_\id) = \tr{(\rho_\id \rho_T)} = \frac{1}{d^2} \sum_\alpha \abs{ \dbraket{\idty}{V_\alpha} }^2 = \frac{1}{d^2} \sum_\alpha \abs{ \tr{V_\alpha} }^2.
\end{equation}
This implies that, for any unitarily implemented channels $\hat{U}$ and $\hat{V}$, we have the relation
\begin{equation}
\cF(\hat{U},\hat{V}) = \cF(\widehat{U^*V},\id),
\end{equation}
where $\widehat{U^*V}$ is the unitarily implemented channel  $\maps{\rho}{U^*V\rho V^*U}$.  In other words, distinguishing between two unitarily implemented channels is equivalent to distinguishing between a suitably chosen unitarily implemented channel and the identity channel.

\section{Fidelity for quantum channels and Uhlmann's theorem}

Our next step is to obtain a meaningful analogue of Uhlmann's theorem for the channel fidelity $\cF$.  In order to do that, we must draw the connection between the channel $T$ and purifications of the density operator $\rho_T$.

First let us prove the following:  {\em given a channel $\map{T}{\cS(\cH)}{\cS(\cK)}$, where $\cH$ and $\cK$ are isomorphic Hilbert spaces $(\cH \simeq \cK)$, the density operator $\rho_T$ is pure if and only if the channel $T$ is unitarily implemented}.  Proving the forward implication is easy:  $\rho_{\hat{U}} = (1/d)\dketbra{U}{U}$, which is a pure state.  Let us now prove the reverse implication.  Suppose that, given the channel $T$, the state $\rho_T$ is pure.  As evident from Eq.~(\ref{eq:roptp}), the reduced density operator $\ptr{\cK}{\rho_T}$ is a multiple of the identity, i.e., it is a maximally mixed state.  Since $\rho_T$ is pure, the reduced density operators $\ptr{\cK}{\rho_T}$ and $\ptr{\cH}{\rho_T}$ have the same nonzero eigenvalues \cite{lindblad}.  All eigenvalues of $\ptr{\cK}{\rho_T} \equiv (1/d)\idty$ are equal and positive.  Since $\cH \simeq \cK$ by assumption, $\ptr{\cK}{\rho_T}$ and $\ptr{\cH}{\rho_T}$ are isospectral.  Hence $\rho_T$ is a maximally entangled state and therefore has the form $(1/d)\dketbra{U}{U}$ for some unitary $U$ (see the paper by D'Ariano et al. in Ref.~\cite{royer} for a proof).  Using Eq.~(\ref{eq:matrel1}), we can write
\begin{equation}
\dketbra{U}{U} = (\tensprod{U}{\idty})\dketbra{\idty}{\idty}(\tensprod{U^*}{\idty}),
\end{equation}
which implies that $T = \hat{U}$, i.e., $T$ is a unitarily implemented channel.

Therefore, for any two unitarily implemented channels $\hat{U}$ and $\hat{V}$, the states $\rho_{\hat{U}},\rho_{\hat{V}}$ are already pure and, as we have shown in the previous section,
\begin{equation}
\cF(\hat{U},\hat{V}) = \frac{1}{d^2} \abs{\dbraket{U}{V}}^2 \equiv \frac{1}{d^2} \abs{\tr{(U^*V)}}^2,
\label{eq:fidunitary}
\end{equation}
which is nothing but the squared normalized Hilbert-Schmidt product of the operators $U$ and $V$.  As we shall now show, the fidelity $\cF(S,T)$ for {\em arbitrary channels} $S$ and $T$ can be expressed as a maximum of expressions similar to the right-hand side of Eq.~(\ref{eq:fidunitary}), but with the difference that in place of the unitaries $U$ and $V$ there will appear certain {\em isometries} from $\cH$ to $\tensprod{\cK}{\cE}$, where $\cE$ is a suitably defined auxiliary Hilbert space.

For this purpose, it is convenient to consider not the channel $T$, but rather the {\em dual channel} $T_*$ \cite{werner2} which converts observables on $\cK$ into observables on $\cH$, i.e., it is a mapping $\map{T_*}{\cB(\cK)}{\cB(\cH)}$, where, e.g., $\cB(\cH)$ is the space of all (bounded) operators on $\cH$.  The channels $T$ and $T_*$ are related by $\tr{\left(T_*(F)\rho\right)} = \tr{\left(FT(\rho)\right)}$, where $F$ is any observable on $\cK$ and $\rho$ is any density operator on $\cH$.  Roughly speaking, the channel $T$ corresponds to the Schr\"odinger picture of quantum dynamics, while the dual channel $T_*$ describes the Heisenberg picture.

We will need to make use of the Stinespring dilation \cite{werner2,sti} of the dual channel $T_*$.  That is, given $T_*$, there exist a Hilbert space $\cE$ and an isometry $\map{V}{\cH}{\tensprod{\cK}{\cE}}$ (i.e., $V^* V = \idty_\cH$) such that, for any $X \in \cB(\cK)$,
\begin{equation}
T_*(X) = V^*(\tensprod{X}{\idty_\cE})V.
\label{eq:stinespring}
\end{equation}
The isometry $V$ contains complete information about the channel $T_*$ (or, for that matter, $T$) and is unique up to a unitary transformation of $\cE$.  In order to arrive at decomposition (\ref{eq:stinespring}) for $T_*$ from its Kraus representation, fix a set $\set{V_\alpha}^n_{\alpha=1}$ of Kraus operators [it is easy to see that if $T(\rho)$ is determined from Eq.~(\ref{eq:krausrep}), then $T_*(X) = \sum_\alpha V^*_\alpha X V_\alpha$]. Let $\cE$ be a Hilbert space of dimension $n$.  We pick an orthonormal basis $\set{\ket{e_\alpha}}^n_{\alpha=1}$ for $\cE$ and define the isometry $V$ through its action on an arbitrary vector $\ket{\psi} \in \cH$:
\begin{equation}
V\ket{\psi} \defeq \sum_\alpha
	\tensprod {V_\alpha \ket{\psi}} {\ket{e_\alpha}}.
\end{equation}
The action of the adjoint operator $V^*$ on product vectors (elementary tensors) $\tensprod{\ket{\psi}}{\ket{\varphi}} \in \tensprod{\cK}{\cE}$ is given by
\begin{equation}
V^* \tensprod {\ket{\psi}} {\ket{\varphi}} 
	\defeq \sum_\alpha \braket{e_\alpha}{\varphi}
	V^*_\alpha \ket{\psi},
\end{equation}
and then extended to all of $\tensprod{\cK}{\cE}$ by linearity.  The fact that $V$ is an isometry can be easily verified; it is also straightforward to confirm Eq.~(\ref{eq:stinespring}). 

Let $\cH,\cK$ be Hilbert spaces with $d = \dim \cH$ and $d' = \dim \cK$. Our goal is to compute the fidelity $\cF(S,T)$ for channels $\map{S,T}{\cH}{\cK}$ by "comparing" various Stinespring dilations of their dual channels.  In order to do that, we will relate purifications of $\rho_S$ and $\rho_T$ to suitable Stinespring dilations of $S_*$ and $T_*$ respectively, and then invoke Uhlmann's theorem for the mixed-state fidelity \cite{jozsa,uhlmann1}.

Since $\rho_T$ is a density operator, it possesses a spectral decomposition
\begin{equation}
\rho_T = \sum^{dd'}_{\alpha=1} v_\alpha \dketbra{\tilde{V}_\alpha}{\tilde{V}_\alpha},
\end{equation}
where $\dbraket{\tilde{V}_\alpha}{\tilde{V}_\beta} = \tr{(\tilde{V}^*_\alpha \tilde{V}_\beta)} = \delta_{\alpha \beta}$.  The operators $V_\alpha \defeq \sqrt{d v_\alpha} \tilde{V}_\alpha$ are then the Kraus operators of $T$ (those $V_\alpha$ that correspond to the zero eigenvalues of $\rho_T$ do not enter into the description of the channel $T$ at all).  The purifications of $\rho_T$ can thus be constructed in a $(dd')^2$-dimensional Hilbert space.  For instance, the state
\begin{equation}
\ket{\varphi_T} \defeq \sum^{dd'}_{\alpha=1} \sqrt{v_\alpha}
	\tensprod {\dket{\tilde{V}_\alpha}} {\ket{e_\alpha}} = \frac{1}{\sqrt{d}} \sum^{dd'}_{\alpha=1}
	\tensprod {\dket{V_\alpha}} {\ket{e_\alpha}},
\label{eq:purifyt}
\end{equation}
where $\set{\ket{e_\alpha}}^{dd'}_{\alpha=1}$ is an orthonormal basis of $\cE \simeq \tensprod{\cK}{\cH}$, is a purification of $\rho_T$ in the Hilbert space $\cE^{\otimes 2}$.  It is quite easy to see that
\begin{equation}
\ket{\varphi_T} = \frac{1}{\sqrt{d}} V_{\rm ext} \dket{\idty},
\end{equation}
where $\map{V_{\rm ext}}{\cH^{\otimes 2}}{\tensprod{\cK}{\cH}\tensprod{}{\cE}}$ is an isometry defined, for an arbitrary $\ket{\psi} \in \cH^{\otimes 2}$, by
\begin{equation}
V_{\rm ext}\ket{\psi} \defeq \sum^{dd'}_{\alpha=1}
	\tensprod
		{(\tensprod{V_\alpha}{\idty}\ket{\psi})}
		{\ket{e_\alpha}}.
\end{equation}
The same thing can be carried out for $\rho_S$.  Upon defining the purification
\begin{equation}
\ket{\varphi_S} \defeq \frac{1}{\sqrt{d}} \sum^{dd'}_{\alpha=1}
	\tensprod {\dket{W_\alpha}} {\ket{f_\alpha}}
\end{equation}
of $\rho_S$, the derivation of $\ket{\varphi_S} = (1/\sqrt{d}) W_{\rm ext}\dket{\idty}$ proceeds along the same lines.

Now, according to Uhlmann's theorem \cite{jozsa,uhlmann1},
\begin{equation}
F(\rho_S,\rho_T) = \max_{\ket{\varphi_S},\ket{\varphi_T}}
	\abs{ \braket{\varphi_S}{\varphi_T} }^2.
\end{equation}
As evident from the discussion above, any purification of $\rho_T$ can be written in the form $(1/\sqrt{d})V_{\rm ext}\dket{\idty}$, where the isometry $V_{\rm ext}$ is determined by the eigenvectors $\dket{\tilde{V}_\alpha}$ of $\rho_T$ (or by their images under any unitary transformation of $\tensprod{\cK}{\cH}$) and by the choice of basis $\set{\ket{e_\alpha}}$ of $\cE$.  The same goes for $\rho_S$, and we can therefore write
\begin{equation}
F(\rho_S,\rho_T) = \frac{1}{d^2} \max_{V_{\rm ext},W_{\rm ext}} 
	\abs{	\dbraket{\idty}{(V_{\rm ext}^* W_{\rm ext}) \idty} }^2.
\end{equation}
It is then an easy exercise to show that
\begin{equation}
\dbraket{\idty}{(V_{\rm ext}^*W_{\rm ext})\idty} = \tr{(V^* W)},
\end{equation}
where $V$ and $W$ are isometries $\cH \rightarrow \tensprod{\cK}{\cE}$ that define Stinespring dilations of $T$ and $S$ respectively, using the same Kraus operators $V_\alpha,W_\alpha$ and the same bases $\set{\ket{e_\alpha}},\set{\ket{f_\alpha}}$ that appear in the corresponding expressions for $V_{\rm ext}$ and $W_{\rm ext}$. We thus arrive at the formula
\begin{equation}
\cF(S,T) = \frac{1}{d^2} \max_{V,W} \abs { \tr{(V^* W)} }^2,
\end{equation}
which can be taken as the statement of {\em Uhlmann's theorem for the channel fidelity $\cF$}.

\section{Applications}

Now that we have defined a fidelity measure for quantum channels, and formulated and proved an analogue of Uhlmann's theorem for it, it is time to ask:  can this measure be at all useful?  In this section, we offer some examples which suggest that the answer to the above question is "yes."

\subsection{Distinguishing channels by superdense coding}

It has by now become conventional wisdom in quantum information theory that "you can't extract more than one bit of classical information out of one qubit" (this assertion can be proved rigorously \cite{yo}).  In other words, there is no measurement that would allow reliable discrimination between symbols from an alphabet of size $M > 2$ if the corresponding message (of $\log M$ bits) is "carried" by a two-state quantum-mechanical system.  However, it is possible for one party to send a 2-bit message to another using a shared entangled state of two qubits in a "superdense coding scheme" \cite{superdense} (this scheme can in fact be extended \cite{werner} to permit reliable discrimination between $d^2$ symbols with the use of entangled states on a tensor product of two $d$-dimensional Hilbert spaces).

As noted by Childs et al. \cite{cpr}, the same strategy can be used to improve distinguishability of quantum channels in the following way.  Suppose we want to distinguish between $d^2$ (possibly time-dependent) Hamiltonians $H_m, m \in \set{1,\ldots,d^2}$ acting on a $d$-dimensional Hilbert space $\cH$.  We prepare the maximally entangled state $(1/\sqrt{d}) \ket{\varphi^+_\cH}$ and then compare the states
\begin{equation}
\ket{\Psi_m} \defeq \frac{1}{\sqrt{d}}(
	\tensprod{e^{-iH_m t/\hbar}}{\idty})\ket{\varphi^+_\cH}, \qquad m = 1, \ldots, d^2.
\end{equation}
The idea is to "stop" the entangled state from evolving at such a time $t_0$ that the quantitiy $\max_{m \neq n} \abs{ \braket{\Psi_m}{\Psi_n} }^2$ $(m,n = 1,\ldots,d^2)$ is minimized (i.e., the pure states $\ket{\Psi_m} \in \cH^{\otimes 2}$ are maximally distinguishable).

We can cast this problem into an equivalent form involving the fidelity measure $\cF$.  We note first that, for each $m$, the projector $\ketbra{\Psi_m}{\Psi_m}$ is nothing but the state $\rho_{\hat{U}_m}$ for the unitarily implemented channel $\hat{U_m}$ with $U_m \defeq e^{-iH_m t/\hbar}$.  That is, $\rho_{\hat{U}_m} = (1/d) \dketbra{e^{-iH_mt/\hbar}}{e^{-iH_mt/\hbar}}$ and
\begin{equation}
\abs{ \braket{\Phi_m}{\Phi_n} }^2 = \frac{1}{d^2} \abs{ \tr{\, e^{-i(H_m-H_n)t/\hbar}} }^2 \equiv \cF(\hat{U}_m,\hat{U}_n).
\end{equation}
If we are to distinguish between the Hamiltonians $H_m$ in fixed time, we are faced with the optimization problem
\begin{equation}
\cF_{\rm opt} = \inf_{t \in [0,T] } \max_{m,n \in \set{1,\ldots,d^2} \atop m \neq n} \cF(\hat{U}_m,\hat{U}_n),
\end{equation}
where $T$ is finite. Provided that the Hamiltonians $H_m(t)$ are sufficiently well-behaved (so that the exponentials $e^{-itH_m}$ are continuous), the maximum of $\cF(\hat{U}_m,\hat{U}_n)$ over all pairs $(m,n)$ with $m,n$ distinct is a lower semicontinuous function, and thus attains its infimum on the compact set $[0,T]$.

This problem can be formulated analogously for arbitrary channels $T_m$, where time dependence can be either continuous (say, via time-dependent Kraus operators) or discrete (e.g., when the channels $T_m$ act on a comparable timescale $\tau$, and we can make measurements only at times $n\tau$, where $n = 0,1,2,\ldots$).   

\subsection{Distinguishing channels with preprocessing}

As we have shown earlier, for any three channels $R,S,T$ [where $\map{S,T}{\cS(\cH)}{\cS(\cK)}$ and $\map{R}{\cS(\cK)}{\cS(\cK')}$, and $\cH,\cK,\cK'$ are Hilbert spaces], we have $\cF(R \circ S, R \circ T) \ge \cF(S,T)$.  In other words, when one aims to distinguish between the channels $S$ and $T$, any postprocessing can only make things worse.

It is pertinent to ask:  what about {\em pre}processing?  Given a pair of channels $\map{S,T}{\cS(\cH)}{\cS(\cK)}$, can we find a Hilbert space $\cK'$ and a channel $\map{R}{\cS(\cK')}{\cS(\cH)}$ such that
\begin{equation}
\cF(S \circ R,S \circ T) \le \cF(S,T)?
\end{equation}

It turns out that in some cases the answer is "yes."  In a recent paper \cite{acin}, \acin demonstrated the following remarkable fact.  Suppose that we have a black box that implements one of two unitaries $U_1,U_2 \in \su(2)$. In order to tell what the black box does (or, equivalently, to distinguish between these unitaries) the optimal strategy consists of preparing the maximally entangled state $(1/\sqrt{2})(\ket{00}+\ket{11})$ (here we use the "computational basis" notation so that, e.g., $\ket{00} \defeq \tensprod{\ket{0}}{\ket{0}})$ and then comparing the states obtained from it by application of the channels $\tensprod{\hat{U}_1}{\id}$ and $\tensprod{\hat{U}_2}{\id}$.  In this case, the measure of distinguishability between $U_1$ and $U_2$ (or between the corresponding channels) is given by $\cF(\hat{U}_1,\hat{U}_2)$.

Now suppose that we have $N$ copies of the black box and can run them in parallel.  If we attempt to compare $U^{\otimes N}_1$ and $U^{\otimes N}_2$ on the basis of the fidelity measure $\cF$ alone, we see that $\cF_N \defeq \cF(\hat{U}^{\otimes N}_1,\hat{U}^{\otimes N}_2) = \cF(\hat{U}_1,\hat{U}_2)^N$ (cf. property {\bf CF5} of the channel fidelity).  If $0 < \cF(\hat{U}_1,\hat{U}_2) < 1$, then the fidelity $\cF_N$ will decrease exponentially with $N$, but it will reach zero only in the limit $N \rightarrow \infty$.  However, as \acin showed in a straightforward argument \cite{acin},  there exist an integer $N_0$ and a state $\ket{\Psi} \in \cH^{\otimes N_0}$, where $\cH$ is a single-qubit Hilbert space, such that $\braket{\Psi}{(U^*_1 U_2)^{\otimes N_0})\Psi} \equiv 0$.  This means that the $N_0$-fold tensor product of the channel $\hat{U}_1$ can be distinguished from the $N_0$-fold tensor product of the channel $\hat{U}_2$ {\em perfectly}!

We can express this result in terms of the preprocessing channel $T_\Psi$ that acts on $\cS(\cH^{\otimes N_0})$ by mapping an arbitrary density operator $\rho$ to the projector $\ketbra{\Psi}{\Psi}$ (i.e., the channel $T_\Psi$ describes the act of preparing the state $\ket{\Psi}$). Thus, for any channel $T$ on $\cS(\cH^{\otimes N})$,
\begin{displaymath}
\rho_{T} = \frac{1}{2^N}
	(\tensprod {T_{\rm odd}} {\id_{\rm even}})(\ketbra{\varphi^+_\cH}{\varphi^+_\cH})^{\otimes N},
\end{displaymath}
and the channel fidelity $\cF(S,T)$ is defined, as before, as the mixed-state fidelity $F(\rho_S,\rho_T)$.  Therefore we obtain the result
\begin{equation}
\cF(\hat{U}_1,\hat{U}_2) > \cF(\hat{U}^{\otimes N}_1,\hat{U}^{\otimes N}_2) > \cF(\hat{U}^{\otimes N_0}_1 \circ T_\Psi,\hat{U}^{\otimes N_0}_2 \circ T_\Psi) \equiv 0.
\end{equation}

It certainly is an interesting and important problem to decide whether (and when) preprocessing can improve distinguishability of arbitrary channels.  For this purpose, it suffices to consider only {\em local} preprocessing since, as shown in the paper by D'Ariano et al. in Ref.~\cite{royer}, the action of an arbitrary channel on the maximally entangled state $\dketbra{\idty}{\idty}$ can be represented as the action of a channel of the form $\tensprod{T}{\id}$. Preprocessing could also involve tensoring $S$ and $T$ with some suitably chosen channel $R$, but we can exclude such "catalyst" channels because $\cF$ is multiplicative with respect to tensor products.  Even with these simplifications, however, the preprocessing problem is still quite challenging.  A first natural step would be to demonstrate equivalence of $\cF$ to some other measure of "distance" between quantum channels.  One such measure is induced by the {\em norm of complete boundedness} (or cb-norm) \cite{werner2,holwer} $\cbnorm{\cdot}$.  That is, $\cbnorm{S-T}$ serves as a measure of "closeness" between the channels $S$ and $T$. For any channel $T$, $\cbnorm{T}=1$, and $\cbnorm{\tensprod{S}{T}} = \cbnorm{S}\cbnorm{T}$. Using properties of the cb-norm \cite{werner2,holwer}, as well as the inequality \cite{fvdg}
\begin{displaymath}
2 - 2 \sqrt{F(\rho,\sigma)} \le \trnorm{\rho-\sigma} \le 2 \sqrt{1-F(\rho,\sigma)},
\end{displaymath}
where $\trnorm{\cdot}$ is the trace norm \cite{note1}, it is easy to prove that
\begin{equation}
2-2 \sqrt{\cF(S,T)} \le \cbnorm{S - T}.
\label{eq:lobound}
\end{equation}

The cb-norm distinguishability measure, however, cannot be used to decide the preprocessing problem for $\cF$ since $\cbnorm{S \circ R - T \circ R} \le \cbnorm{S-T}$ so that, as far as the cb-norm criterion is concerned, any pre- or postprocessing can only make channels less distinguishable.  The reason for this is likely to be the following.  For any map $X$ between operator algebras, the cb-norm is defined as the supremum, over all positive integers $n$, of the usual operator norm $\norm{\tensprod{X}{\id_n}}$, where $\id_n$ is the identity map on $n \times n$ complex matrices.  Thus the cb-norm is a much stronger measure of distinguishability than the channel fidelity $\cF$ because preprocessing (at least in terms of initial-state preparation) is implicitly contained in its definition.  It is therefore important to determine whether a tight upper bound on $\cbnorm{S-T}$ in terms of $\cF(S,T)$ can at all be obtained.  The inequality (\ref{eq:lobound}), on the other hand, can be useful for deriving tight lower bounds on channel capacity.

\subsection{Quantum error-correcting codes}

Our final example involves the problem of characterizing the performance of a quantum error-correcting code (QECC).  We briefly recall the basics of QECC's \cite{qecc}.  The state of a system in a $k$-dimensional Hilbert space $\cH$ is protected by isometrically embedding $\cH$ as a $k$-dimensional subspace $\cK$ (the {\em code}) of an $n$-dimensional Hilbert space $\cH_c$ (the {\em coding space}).  Let the effect of errors be modelled by a channel $\map{T}{\cS(\cH_c)}{\cS(\cH_c)}$.  Then, according to a theorem of Knill and Laflamme \cite{qecc}, the subspace $\cK$ of $\cH_c$ can serve as a QECC for $T$ if and only if there exists a channel $\map{R}{\cS(\cH_c)}{\cS(\cH_c)}$, such that $\left. R \circ T \right|_\cK = \id$, where $\left. T \right|_\cK$ is the restriction of $T$ to $\cK$.  The channel $R$ is called the {\em recovery channel}.

Let $\set{\ket{e_i}}$ be an orthonormal basis of $\cK$, and define the corresponding state $\ket{\varphi^+_\cK}$.  Then the above necessary and sufficient condition is equivalent \cite{qecc} to the requirement that $(R \circ T) (\ketbra{\varphi^+_\cK}{\varphi^+_\cK}) = \ketbra{\varphi^+_\cK}{\varphi^+_\cK}$.  In other words, $\cK$ is a $T$-correcting code if and only if $\cF(\left. R \circ T \right|_\cK, \id) = 1$.

For the special case of comparing a channel $T$ with the identity channel, we can derive upper and lower bounds on $\cF(T,\id)$ in terms of $\cbnorm{T-\id}$.  For this purpose we need the {\em off-diagonal fidelity} of the channel $\map{T}{\cS(\cH)}{\cS(\cH)}$, defined by \cite{werner2}
\begin{equation}
\cF_\%(T) \defeq \sup_{\ket{\psi},\ket{\varphi} \in \cH}
	\rp {\braket {\varphi} {T(\ketbra{\varphi}{\psi})\psi} },
\end{equation}
for which we have the inequality \cite{werner2}
\begin{equation}
\cbnorm{T-\id} \le 4 \sqrt{1-\cF_\%(T)}.
\end{equation}
Then $\cF(T,\id) \le \cF_\%(\tensprod{T}{\id})$, so that, using the fact that the cb-norm is multiplicative with respect to tensor products, we get
\begin{equation}
\cbnorm{T-\id} \le 4\sqrt{1-\cF(T,\id)}.
\label{eq:upbound}
\end{equation}
Combining inequalities (\ref{eq:lobound}) and (\ref{eq:upbound}) yields
\begin{equation}
\left(1-\frac{1}{2}\cbnorm{T-\id}\right)^2 \le \cF(T,\id) \le 1 - \frac{1}{16}\cbnorm{T-\id}^2.
\end{equation}
The upper bound in this inequality is not nearly as tight as the lower bound.  Indeed, when $\cbnorm{T-\id}$ equals its maximum value of 2, the fidelity $\cF(T,\id)$ can take any value between 0 and 3/4.  This serves as yet another indication that the cb-norm is a much more stringent distinguishability criterion than the channel fidelity $\cF$.  

\section{Conclusion}

In this letter, we have proposed a fidelity measure $\cF$ for quantum channels.  This fidelity measure possesses properties that are similar to the properties of the mixed-state fidelity $F$. We have stated and proved an analogue of Uhlmann's theorem for $\cF$ and discussed possible applications of this fidelity measure to problems in quantum information science.  We have also outlined the way $\cF$ is related to another criterion of channel distinguishability, the cb-norm distance.

\section*{Acknowledgments}

This work was supported by the U.S. Army Research Office through MURI grant DAAD19-00-1-0177.


\begin{thebibliography}{}
\bibitem{note1}For any pair $\rho,\sigma$ of density operators, the trace norm $\trnorm{\rho-\sigma}$ is twice the trace distance between $\rho$ and $\sigma$ \cite{nielchu}.
\end{thebibliography}
\end{document}